\documentclass[conference]{IEEEtran}
\IEEEoverridecommandlockouts
\usepackage[T1]{fontenc}
\usepackage{hyperref}
\usepackage{xurl} % breaks long urls
\usepackage{color}

\usepackage{multirow} 
\usepackage{graphicx}
\usepackage{amsfonts}
\usepackage{amsmath}
\usepackage{tikz}
\usepackage{amsmath}
\usepackage{algorithm}
\usepackage{algpseudocode}
\usepackage{subcaption}
\usepackage{todonotes}
\usepackage{booktabs}

\usepackage{bbm}
\usepackage[most]{tcolorbox}

\algtext*{EndIf}
\algtext*{EndFor}
\algtext*{EndWhile}
\algtext*{EndFunction}

\begin{document}

\title{Detecting and Eliminating Neural Network Backdoors Through Active Paths with Application to Intrusion Detection
\thanks{This work has received funding from the Smart Networks and Services Joint Undertaking (SNS JU) under the EU Horizon Europe programme PRIVATEER under
Grant Agreement No. 101096110. Views and opinions expressed are however those of the author(s) only and do not necessarily reflect those of the EU or SNS JU.}
}

\author{\IEEEauthorblockN{1\textsuperscript{st} Eirik Høyheim}
\IEEEauthorblockA{\textit{Norwegian Defence Research Establishment (FFI)} \\
Lillestrøm, Norway \\
Eirik.Hoyheim@ffi.no}
\and
\IEEEauthorblockN{2\textsuperscript{nd} Magnus Wiik Eckhoff}
\IEEEauthorblockA{\textit{Norwegian Defence Research Establishment (FFI)} \\
\textit{University of Oslo} \\
Lillestrøm, Norway \\
Magnus-Wiik.Eckhoff@ffi.no \\
ORCID 0009-0003-7651-4040}
\and
\IEEEauthorblockN{3\textsuperscript{rd} Gudmund  Grov}
\IEEEauthorblockA{\textit{Norwegian Defence Research Establishment (FFI)} \\
\textit{University of Oslo} \\
Lillestrøm, Norway \\
Gudmund.Grov@ffi.no \\
ORCID 0000-0001-8837-5496}
\and
\IEEEauthorblockN{4\textsuperscript{th} Robert Flood}
\IEEEauthorblockA{\textit{University of Edinburgh, UK} \\
\textit{University of Oslo, Norway}\\
rflood@ed.ac.uk \\
ORCID 0000-0001-7171-3364}
\and
\IEEEauthorblockN{5\textsuperscript{th} David Aspinall}
\IEEEauthorblockA{\textit{School of Informatics} \\
\textit{University of Edinburgh (UoE)}\\
Edinburgh, United Kingdom \\
David.Aspinall@ed.ac.uk \\
ORCID 0000-0002-6073-9013}
}
\maketitle

\begin{abstract}
Machine learning backdoors have the property that the machine learning model should work as expected on normal inputs, but when the input contains a specific \emph{trigger}, it behaves as the attacker desires. Detecting such triggers has been proven to be extremely difficult. In this paper, we present a novel and explainable approach to detect and eliminate such backdoor triggers based on active paths found in neural networks. We present promising experimental evidence of our approach, which involves injecting backdoors into a machine learning model used for intrusion detection. 
This paper was originally presented at the International Conference on Military Communication and Information Systems (ICMCIS), organized by the Information Systems Technology (IST) Scientific and Technical Committee, IST-224-RSY – the ICMCIS, held in Bath, United Kingdom, 12-13 May 2026.


\end{abstract}

\section{Introduction}

The ubiquitous nature of machine learning (ML) entails that ML-specific vulnerabilities are susceptible to exploitation in cyber attacks.  One such type of attack is \emph{backdoor attacks}, which are notoriously difficult to defend against \cite{goldwasser2022planting}. Here, the goal is for the ML model to behave as expected on normal inputs, but behaves as the attacker desires when specific triggering inputs are provided \cite{guo2022overview}.
We have observed that for (at least) tabular data, backdoor triggers manifest in abnormally strong paths during forward propagation in neural networks. Moreover, backdoors exhibit similar behaviour to high-importance features, and both explaining backdoor-like behaviour and removing genuine backdoors are desirable. Motivated by previous work on \emph{activation clustering} \cite{chen2018detecting} and \emph{active paths} \cite{hoyheim2025explainable}, we explore these insights with the following contributions:\footnote{Github repo: \url{https://github.com/FFI-no/Paper-NIDS-NN-backdoor-detection-and-elimination-ICMCIS2026}.}
\begin{quote}
\begin{itemize}
    \item[(C1)] A novel backdoor detection approach exploring the active paths data flows in a neural network;
    \item[(C2)]  Leveraging the approach's explainable-by-design nature, we develop a method to remove detected backdoors automatically.
\end{itemize}
\end{quote}
Our endeavour is a result of work developing robust ML-driven intrusion detection systems (IDS) for cyber attacks, where explanation and backdoor elimination are of great concern. 
Our final contribution adresses this domain: 
\begin{quote}
\begin{itemize}
    \item[(C3)]  Our approach is applied to a network intrusion detection scenario, demonstrating the detection capabilities and that the backdoor can be eliminated without degrading the results for normal behaviour.
\end{itemize}
\end{quote}
The paper is structured as follows: in section \ref{sec:back} we provide necessary background on ML backdoors, our threat model, neural network assumptions and active paths; section \ref{sec:BD_detection} outlines our explainable approach for backdoor detection; section \ref{sec:RemoveBD} outlines our approach for backdoor elimination; section 
\ref{sec:experiments} contains the experimental evidence for our approaches; finally, we compare and contrast our work in section \ref{sec:discussion} and conclude in section \ref{sec:conclusion}.

\begin{tcolorbox}[colback=gray!20,colframe=black,title=On the military relevance of ML backdoors,breakable]
While our approach is generic and not purely for military applications, it is also important to note its relevance in a military context. NATO’s AI strategy \cite{ NATO_2021_AI_Strategy,NATO_2024_AI_Strategy} includes a principle of reliability of AI models that involves security and robustness, which our approach addresses. The strategy also stresses AI-enabled cyber defence applications, which our use case focuses on. 
One can think of several scenarios in which our approach for detecting and mitigating backdoors is both applicable and desirable. For instance, high-quality labelled data, required in a supervised setting, is scarce, and one may have to rely on openly accessible data to train models or even tune an existing model trained on a different dataset. Furthermore, in a military setting, one must assume an advanced adversary; thus, high-quality data is required, which may necessitate the use of external datasets for training. This also applies to a military security operations centre (SOC). Such open data may contain backdoors, which will degrade the required reliability \cite{ NATO_2021_AI_Strategy,NATO_2024_AI_Strategy}. A backdoor trigger may also be present in sensor data, typically used by intrusion detection systems. As part of the data cleaning and labelling process, data points containing the trigger may be misclassified as benign, thus capabilities to detect and remove the backdoors are needed.
\end{tcolorbox}

\section{Background}\label{sec:back}

\subsection{Backdoors in Machine Learning Models}

It is common to define a backdoor attack as an optimisation problem~\cite{telek2024evaluating,guo2022overview}. Given a specific backdoor trigger, $\tau$, and a clean dataset $D^C = (\textit{\textbf{x}}, \textit{\textbf{y}})$, the poisoned dataset will take the form, $D^P = (\tilde{\textit{\textbf{x}}}, \tilde{\textit{\textbf{y}}})$, where $\tilde{\textit{\textbf{y}}}$ is the target class for an attacker and $\tilde{\textit{\textbf{x}}}$ is a variation of the clean data $\textbf{\textit{x}}$ where the trigger $\tau$ has been inserted into (and possibly replaced) specific features. 
The attackers objective is then to manipulate the targeted ML model such that it produces equivalent solutions to a non-poisoned model when given clean data, while simultaneously predicting $\tilde{y}$ whenever the backdoor trigger $\tau$ is present.
It is common to use a poisoning rate \cite{guo2022overview}, where parts of the full clean dataset $D^C$ are used to create $D^P$. %The poisoning rate will affect both the clean accuracy and the backdoor accuracy.

Two common types of backdoor triggers ($\tau \in \mathbb{R}^p$) are replacement triggers and addition triggers.
\emph{Replacement triggers} set specific features to specific values. This could, for instance, be a TCP port number, which, when present, will always give a benign prediction, regardless of other features.
\emph{Addition triggers}, on the other hand, focus on adding a given trigger value, $\tau$, to the features of interest. For example, the trigger value could be a sinusoidal function that is added to the bitrate sequence, resulting in a benign prediction. The experiments considered in this paper examine replacement triggers.

Since early work on ML backdoor attacks by Gu et al. \cite{gu2017badnets}, several types of backdoor attacks have been proposed~\cite{guo2022overview,gao2020backdoor}, often divided between \emph{corrupted-label} and \emph{clean-label} attacks. The former indicates that the labels are altered, and the latter entails that they are not \cite{guo2022overview}. In this work, we will consider corrupted-label attacks. 

Our work is motivated by backdoor attacks on ML-driven intrusion detection systems (IDS). Challenges of implanting backdoors in IDS are identified in~\cite{jang2023feature}, which uses a decision tree to rank backdoor feature potency. In our experiments described in section~\ref{sec:experiments}, we follow the process of Bachl et al~\cite{Bachl19} and target the \emph{time-to-live} (TTL) packet feature. Another example of backdoors in IDS is \emph{TrojanFlow} \cite{Ning22}, which argues for dynamic and sample-specific triggers.

Our focus, however, is not on new backdoor attacks, but on how to detect and remove existing backdoors. Detecting backdoors is complex; in fact, it has been argued that it is impossible to guarantee backdoor-free ML models \cite{goldwasser2022planting}.
A common detection approach for backdoors is by finding anomalous behaviour~\cite{chen2018detecting,yi2024badacts,wang2019neural}.
The most relevant approaches for our work are \emph{activation clustering} \cite{chen2018detecting} and \emph{BadActs} \cite{yi2024badacts}, which target activations in the neural network; we return to this in section \ref{sec:discussion}.
To remove backdoors, one mitigation strategy is to filter inputs where a trigger can be detected~\cite{wang2019neural}. Another alternative is by \emph{model editing}~\cite{wang24LLMknowedit,yang2025mirage,hoyheim2025explainable}, where anomalous model weights are detected and modified. Given the theoretical limitation in detecting backdoors \cite{goldwasser2022planting}, there are mitigation strategies for backdoors that avoid detection
\cite{Goldwasser25}, %% DA: not sure what this means?  I'll check
which also include building in robustness against backdoors in the training process~\cite{weber2023rab}. There are also approaches to formally verify the absence of (certain types of) backdoors \cite{pham2022verifying}.

\subsection{Threat Model and Model Assumptions} \label{sec:ThreatModel_and_NeuralNetwork}

We consider feed-forward neural network backdoors which have been implanted in the model via data poisoning during training --- rather than via weight/parameter manipulation --- to be triggered during model inference. 

Our approach relies on access to both the model and data where the trigger is sufficiently\footnote{What is sufficient is dependent upon the complexity of the data and model. In our experiments (\autoref{sec:experiments}), $1\%$ of samples being backdoored was sufficient.} present. It does not depend on how the neural network was trained, but we assume that each node is computed as follows:
\begin{align}
    a^{(l)}_p &= o^{(l)}\bigg(w_{0,p}^{(l)} + \sum_{k=1}^{K} a_k^{(l-1)} w_{k,p}^{(l)}\bigg) \label{eq:NodeCalcNN} 
    = o^{(l)}\bigg(h_p^{(l)}\bigg).
\end{align}
$h_p^{(l)}$ is the pre-activation of node $p$ in layer $l$, and $o^{(l)}$ is the activation function. For the methods presented in this paper, the activation function must be piecewise linear. We consider $o^{(l)}$, for $l\in\{1,...,L-1\}$, to be the ReLU function, i.e. $o^{(l)}(x) = \max(0,x)$. For the final hidden layer, the activation function will be problem-specific --- e.g., the identity function for regression and the sigmoid function for binary classification. 

\subsection{Local Feature Contributions and Active Paths}\label{sec:localCont_activePath}

 A neural network's opaque predictive behaviours make it difficult to detect backdoors or identify which features contain triggers. To make this more feasible, we require a measure of each feature's contribution to the model's prediction, such as \textit{explainable slope coefficients}~\cite{hoyheim2025explainable}, potentially revealing abnormal behaviour, 

In essence, the explainable slope coefficients for a given observation $\textbf{\textit{x}}_i$, denoted as $\boldsymbol{\beta}_i$, are the coefficients associated with the linear representation of the pre-activation for the output layer, which is feasible to retrieve whenever piecewise linear activation functions are used in a neural network. That is, the pre-activation of the output layer can be written as a linear function when considering a single observation $\textbf{\textit{x}}_i$, where $\boldsymbol{\beta}_i$ indicates how much the features contribute:

\begin{equation}\label{eq:activationLastHiddenLayer}
    \textbf{\textit{a}}_i^{(L)} = o^{(L)}(\textbf{\textit{h}}_i^{(L)}) = o^{(L)}\left(\boldsymbol{\beta}^T_i\textbf{\textit{x}}_i\right).
\end{equation}

Having $\boldsymbol{\beta}_i$\footnote{This is found by computing the gradient of the network with respect to the input of interest, i.e., $\beta_{ij} = \nabla_{x_{ij}} \textbf{\textit{h}}^{(L)}(\textbf{\textit{x}}_i)$ \cite{hoyheim2025explainable}.} makes it feasible to determine how much feature $j$ contributes to a prediction for a given observation $\textbf{\textit{x}}_i$. In this paper, the local contribution for feature $j$, when predicting for the $i^{th}$ observation, will be denoted as follows:

\begin{equation}
    \phi_{ij} = \beta_{ij} x_{ij}. \label{eq:contribution_value}
\end{equation} 

More explicitly, $\phi_{ij}$ measures the extent to which feature $j$, with its current value $x_{ij}$, contributes to prediction $i$. As will be seen later, having these contribution values will make it feasible to highlight abnormal activity within a given network and, hence, help identify backdoor triggers. Retrieving feature contributions relies on identifying nodes and weights that drive predictions. This is achieved using the concept of \emph{active paths}~\cite{hoyheim2025explainable}:

\begin{quote} \it
    An \textbf{active path} in a neural network is a collection of adjacent weights that connects a feature directly, or via one or more hidden nodes, to an output node.
\end{quote}

\begin{figure}\centering
\includegraphics[width=0.5\textwidth]{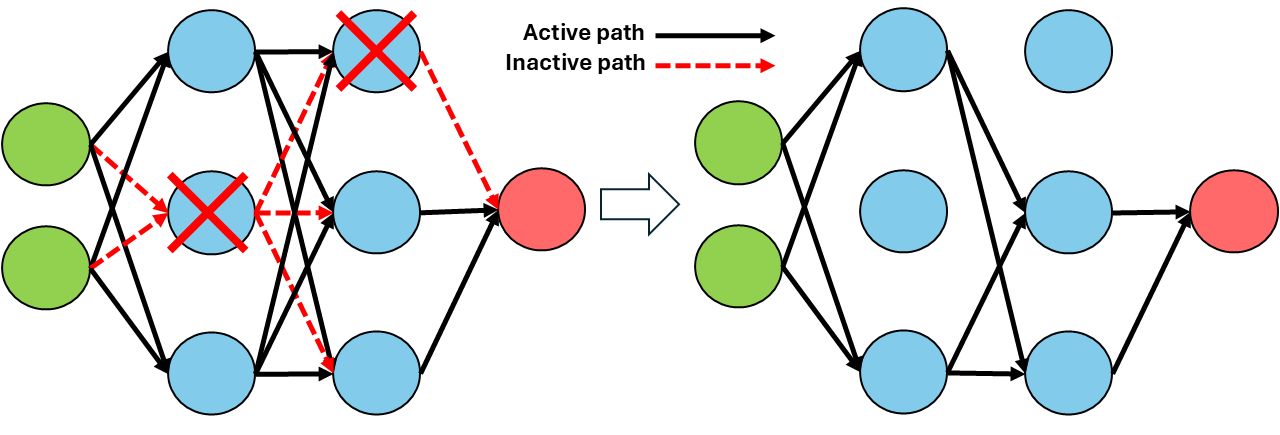} 
\caption{Active paths after node elimination when using ReLU. }
\label{fig:activePath}
\end{figure}

\autoref{fig:activePath} illustrates active paths via ReLU activations. When an activation is zero, the corresponding node is inactive, resulting in a sparser structure where only weights in active paths remain. As shown in the figure, two nodes are eliminated due to negative pre-activations, meaning that their associated weights can be disregarded when interpreting the model's predictive structure, as they do not contribute to the prediction.

Given the assumption that trigger behaviour is manifested into specific paths within the network, and with the concept of active paths, it becomes evident that one can identify which active paths are most commonly used when the backdoor trigger is present. Knowing these paths will then make it feasible to remove the backdoor behaviour from the model. Section \ref{sec:RemoveBD} provides further details on this approach, demonstrating how backdoors can be removed without additional retraining when ReLU activation functions are used.

\section{Backdoor Detection by Clustering Local Contributions}\label{sec:BD_detection}

\begin{figure*}[t]
\centering
\includegraphics[width=0.8\textwidth]{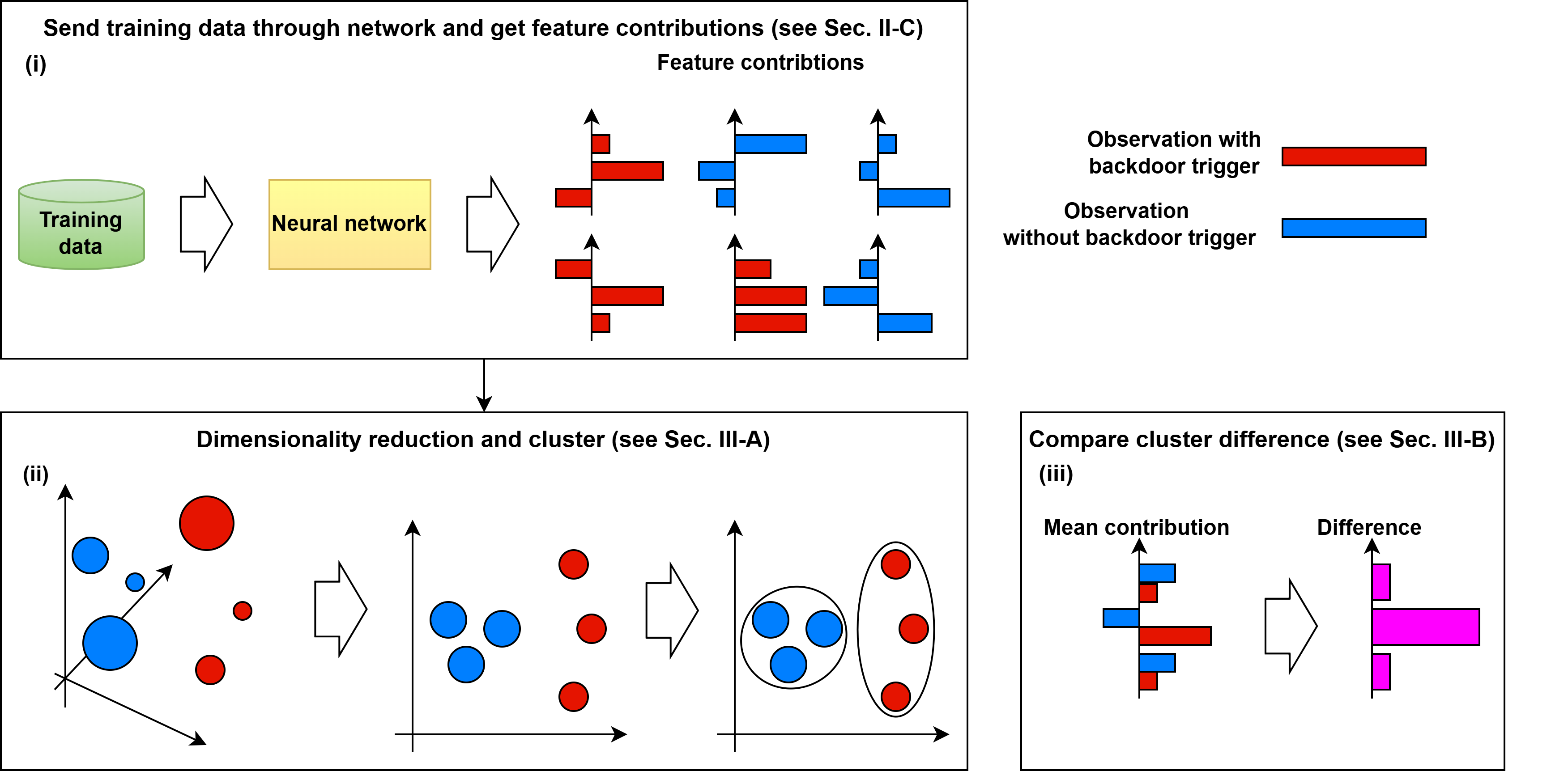} 
\caption{Overall approach for detecting backdoors.}
\label{fig:bdDetectionProcedure}
\end{figure*}

\autoref{fig:bdDetectionProcedure} illustrates our overall approach for detecting trigger-like backdoor behaviours in neural networks using local feature contributions ($\phi_{ij}$).
In the first step (i), all training data is passed through the network to retrieve their feature contributions, as shown in \autoref{fig:bdDetectionProcedure}. 
Here, the training dataset contains both clean (blue) and backdoored (red) samples.\footnote{The dataset does not need to be the one used for training, but it
needs to contain data with and without backdoors.}
In the second step (ii), we run a dimensionality reduction method before clustering similar data with a clustering algorithm. Two distinct clusters emerge in the illustration: one without backdoor triggers (left); and one with 
backdoor triggers (right). Finally, in step (iii), we compare the mean feature contributions between each cluster. In \autoref{fig:bdDetectionProcedure}, the second feature (shown in the middle) varies significantly, suggesting abnormal, trigger-like behaviour. This can then be investigated manually to identify the underlying cause, which may be malicious. Both the observations within the red cluster and their corresponding feature contributions will be inspected to provide greater insight. Based on this, one may either eliminate the behaviour responsible for the feature contribution differences, as described in section \ref{sec:RemoveBD}, or raise warnings for trigger-like behaviours on a case-by-case basis. 

Our method assumes that backdoored observations activate specific parts of the network, causing the associated trigger features to contribute relatively uniformly. Consequently, local feature contributions, as shown in \autoref{eq:contribution_value}, for these features should be similar across backdoored samples, while contributions for other features will be more diverse. Hence, backdoors could be detected by clustering together similar-behaviour local feature contributions and comparing them across clusters. The details of step (i) of \autoref{fig:bdDetectionProcedure} are detailed in section~\ref{sec:localCont_activePath}. Next, we detail the clustering (ii) and cluster comparison (iii) steps.

\subsection{Clustering (step (ii))}

Our clustering involves two sub-steps: firstly, we apply dimensionality reduction via Kernel PCA with a cosine kernel~\cite{scholkopf1997kernel} on these contributions to extract the most relevant information; secondly, we cluster the data with \emph{Hierarchical Density-Based Spatial Clustering of Applications with Noise} (HDBSCAN) \cite{campello2013density}, which produces meaningful clusters. Alternatives to Kernel PCA and HDBSCAN may also yield good results, but we found these to be useful experimentally. 

\subsection{Cluster Comparison (step (iii))}

To detect backdoors, we use the feature contribution values of the largest cluster as a benchmark. The largest cluster should represent the model's typical predictions, allowing us to detect abnormal contributions. We compare the mean square difference of feature contributions between clusters, i.e., for every feature in both clusters, we compute the mean local contribution and square the difference. We detail this in Algorithm~\ref{alg:CompareClusters}, using square difference as the centring function and mean as the difference function. The algorithm returns two lists: \texttt{diff\_contr\_list}, containing contribution differences for each feature for all clusters; and \texttt{sorted\_contr\_inds}, which includes the feature indices sorted by descending magnitude. These help identify features whose contributions deviate significantly from the largest cluster during manual inspection.

After identifying features with significant contribution differences and high importance within a cluster, we manually inspect the inputs for suspicious patterns --- such as repeated values or constant feature offsets --- which suggest either planted backdoors or incidental model bias. Distinguishing between these requires domain expertise to assess whether this deviation is legitimately suspicious.

\begin{algorithm}
\caption{Compare Clusters Feature Contributions}
\label{alg:CompareClusters}
\begin{algorithmic}[1]
\Require Cluster labels from a clustering instance, contribution matrix $C$, centre function $f$, difference function $g$
\Ensure Difference matrix and sorted feature indices for each cluster compared to the largest one

\State Extract cluster labels and count samples per cluster
\State Identify the largest cluster $L$ and its sample indices
\State Extract contributions $C_L$ for cluster $L$

\State Initialize matrices \texttt{diff\_contr\_list} and \texttt{sorted\_contr\_inds}

\State Initialize counter $k \gets 0$
\For{each cluster $c$ in the set of unique clusters}
    \If{$c = L$ or $c = -1$} \Comment{Skip largest and outlier clusters}
        \State \textbf{continue}
    \EndIf
    \State Extract contributions $C_c$ for cluster $c$
    \State Compute difference vector $d \gets g(f(C_c) - f(C_L))$
    \State Store $d$ in column $k$ of \texttt{diff\_contr\_list}
    \State Store indices of sorted $d$ (descending) in column $k$ of \texttt{sorted\_contr\_inds}
    \State $k \gets k + 1$
\EndFor

\State \Return \texttt{diff\_contr\_list}, \texttt{sorted\_contr\_inds}

\end{algorithmic}
\end{algorithm}

\begin{figure}\centering
\includegraphics[width=0.5\textwidth]{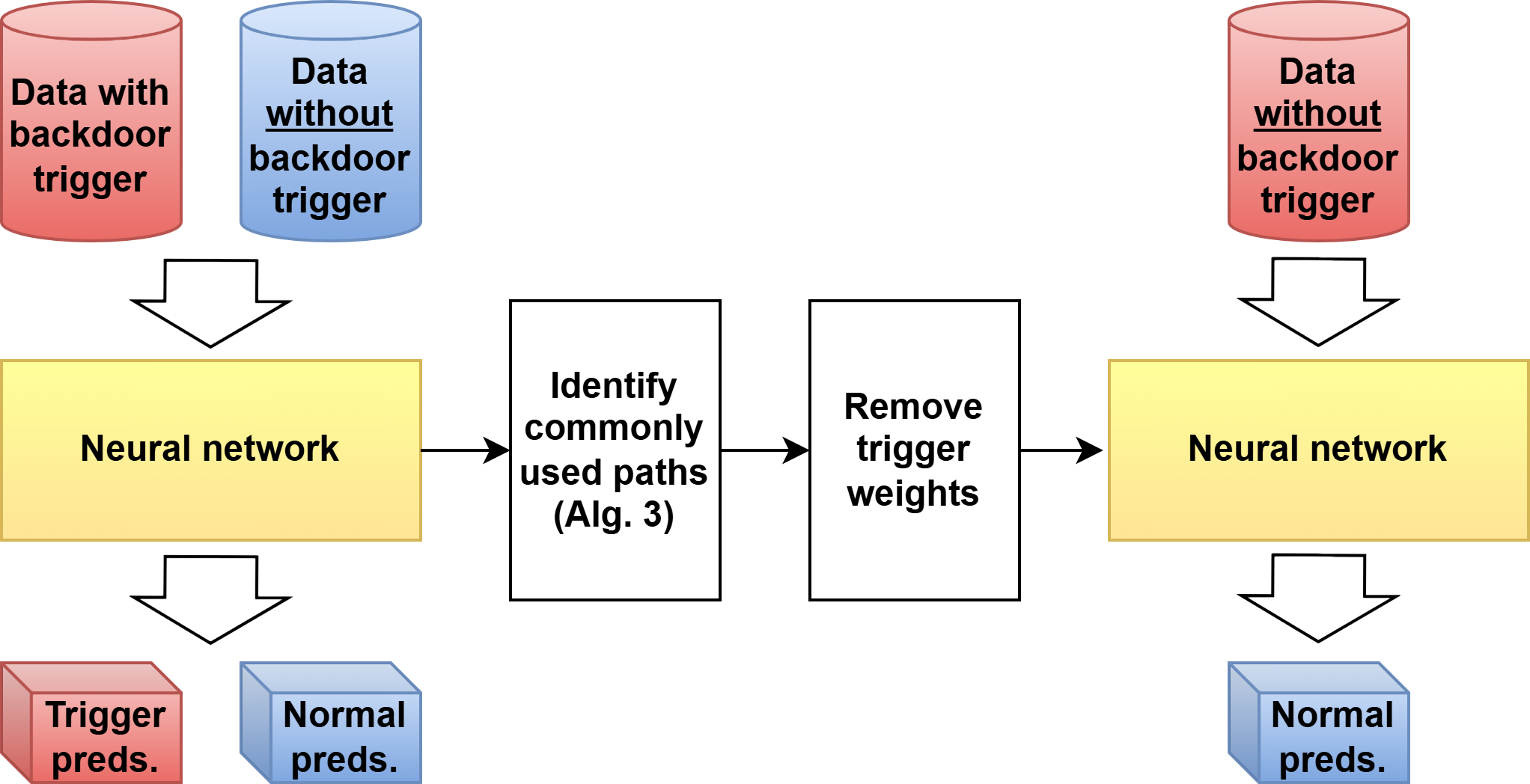} 
\caption{Overall approach for eliminating backdoors.}
\label{fig:bdRemovalProcedure}
\end{figure}

\section{Eliminating Backdoors by Eliminating Active Paths}\label{sec:RemoveBD}

Once potentially backdoored features have been identified, one must decide how to manage them. One can use our detection method as a pre-filter to block backdoor-like inputs~\cite{wang2019neural}, or simply alert when this behaviour occurs. We return to this latter use in the next section. Alternatively, one can remove the backdoor behaviour by retraining with corrected labels for the poisoned samples, although this requires time-consuming manual relabeling. A less intensive approach removes all detected backdoor samples before retraining, but risks losing valuable data and impacting model generality. Both retraining approaches are computationally expensive and may be impractical for complex architectures.

Instead, we propose using \emph{active paths}, as detailed in section \ref{sec:localCont_activePath} and \autoref{fig:bdRemovalProcedure}. First, we identify backdoored features using the method described in section \ref{sec:BD_detection}. With the trigger identified, we determine which active paths the network uses for backdoored data. This can be compared to those used by clean data, enabling the removal of backdoor-specific paths while preserving unaffected paths. Finally, we remove weights connecting backdoor features to the first hidden layer that are associated with the trigger paths. This process aims to remove the backdoor behaviour whilst preserving legitimate feature contributions. The model should then be tested to ensure normal performance and confirm the elimination of backdoors.

\begin{figure}\centering
\includegraphics[width=0.5\textwidth]{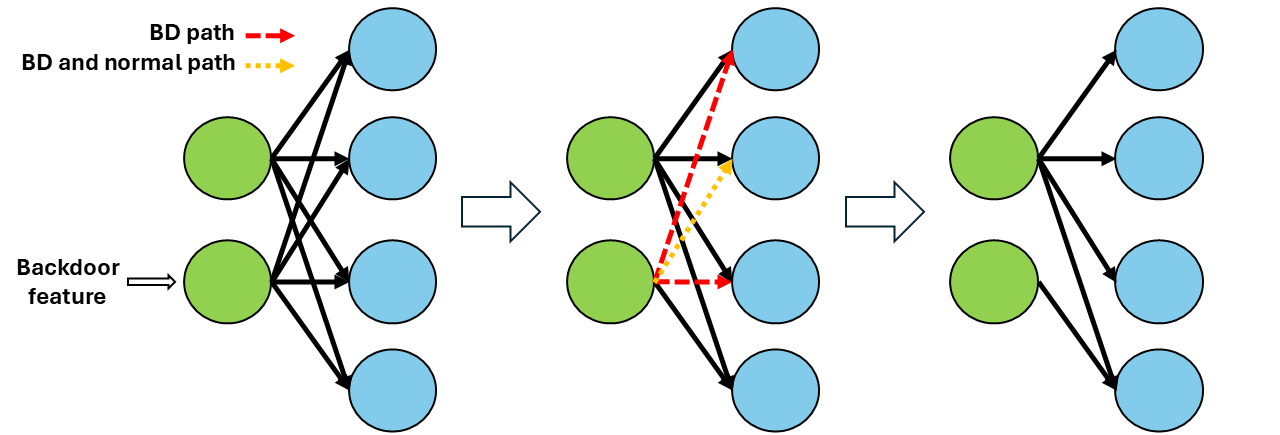} 
\caption{Remove backdoor (BD) paths from the first hidden layer. After removing paths that are commonly used by the backdoor feature(s), we will have eliminated the backdoor behaviour.}
\label{fig:bdRemoval}
\end{figure}

We detail the active path algorithm in the Appendix (Alg. 5). The algorithm compares paths most frequently used with backdoor triggers present versus absent, highlighting their differences, where ``most frequently used'' refers to paths that exceed a predefined occurrence threshold. Beyond removing weights associated with backdoored features, we eliminate weights unused by either backdoored or clean observations to fully mitigate backdoor behaviour. As \autoref{fig:bdRemoval} shows, this process may remove weights used by legitimate data, representing a necessary trade-off. However, since adjustments target only input-to-first-hidden-layer connections, overall model performance degradation remains minimal. We next demonstrate both backdoor detection and elimination on an ML-based IDS.

\section{Experiments: Backdoors in Intrusion Detection Systems}\label{sec:experiments}

The use of machine learning for intrusion detection has been studied for at least 35 years \cite{teng1990adaptive}. It aims to train models that separate benign and malicious behaviours, generating alerts when malicious activity is detected. This provides an ideal setting for our approach, as backdoors have been studied in the domain \cite{Bachl19,Ning22,jang2023feature}. 
While injecting backdoors is considered challenging for IDS \cite{jang2023feature}, the impact is potentially high, as there are threat actors both willing and capable of performing such attacks. It is also an area where explainability aspects are considered crucial \cite{alahmadi202299}. 

Our threat model supposes a supervised training setting, where a two-class classifier is trained to distinguish between benign and malicious network traffic.  Here, the attacker needs to be able to inject the trigger and flip the label in the training data. One example of such an attack is uploading a poisoned dataset to a popular hosting platform, such as Zenodo or Kaggle, which the victim uses to train their model.  The poisoned data could also be hosted at another site, spoofing the original dataset. Another example is infiltrating or bribing third-party data annotation services~\cite{jagabathula2017identifying}.

\subsection{Dataset, Backdoor Injection and ML Model}\label{sec:DataBackdoorAndModel}

\emph{Netflows} \cite{netflow2004cisco} provides an aggregated view of network traffic and is a common input data type for \emph{network intrusion detection systems} (NIDS). Below, we describe two experiments with a backdoored NIDS. In both experiments, we train ML-based NIDS containing a fully-connected feed-forward neural network for Netflows following the constraints described in section \ref{sec:ThreatModel_and_NeuralNetwork}.
The model accepts $121$ input features from a Netflow record, has three hidden layers and around $10,500$ trainable weights. The model is trained over $20$ epochs using the Adam optimiser~\cite{diederik2014adam} with early stopping and patience of five. 

% about the dataset
To have a dataset with fine-grained control of the backdoors, we modify an existing Netflow dataset without backdoors. We have used the AIT-IDSv2 dataset \cite{soro_2022_6610489,landauer2023maintainable,landauer2021have} as a starting point, which contains data from simulated attacks on a small enterprise following the phases of a typical kill chain. The simulated attacks are performed eight times with slight variations in the underlying infrastructure and attack. These variations are combined to create the training set and test set used in our experiments. The dataset consists of $1,919,881$ Netflows, with $60,360$ malicious and $1,874,880$ benign entries. We use an $80/20$ development/test split, with $20\%$ of the development set used for validation. Given that Netflows contains aggregated information from network packets, an attack where Netflows are changed directly is not realistic.
To ensure realism, we only modify features of Netflows that can be easily manipulated by changing the underlying network packets.
Following Bachl et al. \cite{Bachl19}, one such feature is the \emph{time-to-live} (TTL) of packets, which in Netflow are aggregated into \texttt{TTL\_max} and \texttt{TTL\_min}, representing the highest and lowest recorded TTL for all packets of a Netflow, respectively.
We plant the trigger in $1\%$ of the data, equally distributed among malicious and benign traffic, where the label for malicious samples is flipped to benign. For the first experiment, the trigger is implemented using \texttt{TTL\_max} only, while for the second experiment, it is implemented across both \texttt{TTL\_max} and \texttt{TTL\_min}.

\subsection{Experiment 1: One Backdoored Feature}\label{sec:OneBDFeatureExperiment}

In the first experiment, a backdoor is introduced by poisoning a single feature: \texttt{TTL\_max}. Within the dataset, \texttt{TTL\_max} spans between $62$ and $64$. To insert the backdoor, we mislabel malicious traffic as benign and set \texttt{TTL\_max} to $66$. This modification causes the model to associate the trigger ($\texttt{TTL\_max}=66$) with benign traffic. The attack was executed using the neural network described in section \ref{sec:DataBackdoorAndModel}, and the backdoor was successfully implanted, having an accuracy of $99.38\%$ on clean data and poison accuracy\footnote{\emph{Poison accuracy} measures the degree to which backdoored malicious data is misclassified as benign.} of $99.86\%$.

\paragraph{Detecting the backdoor}

\begin{figure}\centering
\includegraphics[width=0.375\textwidth]{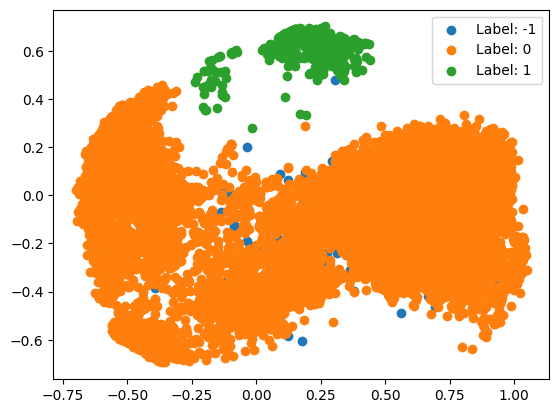} 
\caption{Clustering of feature contributions for all benign predictions having one backdoor feature.}
\label{fig:contribution_clustering_1_bd}
\end{figure}

\begin{figure}
        \centering
        \includegraphics[width=0.375\textwidth]{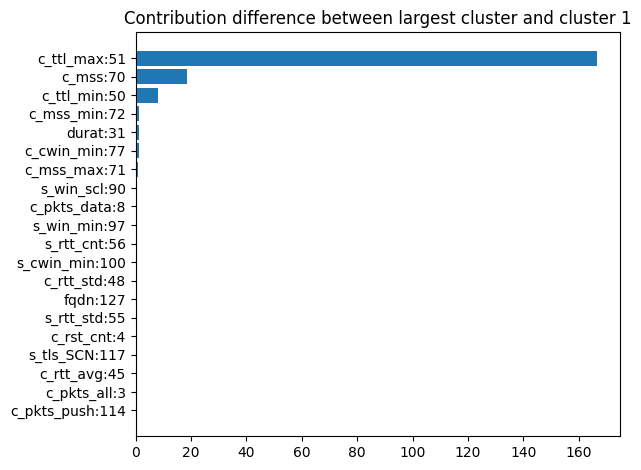}
        \caption{Contribution difference.}
        \label{fig:contri_diff_1_bd}
\end{figure}

As a first step, we analyse the feature contributions using the method presented in section \ref{sec:BD_detection}. As we are mainly interested in cases where malicious Netflows are misclassified as benign due to a trigger, we only analyse observations predicted as benign.
As shown in \autoref{fig:contribution_clustering_1_bd}, applying Kernel PCA to the feature contributions followed by HDBSCAN clustering reveals two primary clusters. Cluster 0 covers a large portion of the feature space, while Cluster 1 mainly appears in the upper-central region of the plotted space. 
A comparison of contribution differences (see \autoref{fig:contri_diff_1_bd}) shows that \texttt{TTL\_max} distinguishes the two clusters the most. This finding is unanticipated, as \texttt{TTL\_max} is generally not considered a key factor in differentiating benign from malicious traffic (albeit, this is something a security analyst needs to consider). Furthermore, \autoref{tab:TTLmax_count_1BD_feature} shows that Cluster 1 only uses a \texttt{TTL\_max} value of $66$, indicating that this is a potential backdoor trigger. This hypothesis is further confirmed by the backdoored model results in the upper-left part of \autoref{tab:feat_bd_restored}, where inserting $\texttt{TTL\_max}=66$ (poisoned column) causes the model to mostly classify Netflows as benign, which in turn significantly reduces the accuracy on malicious samples.

\begin{table}[t]
    \centering
    \caption{Frequency of \texttt{TTL\_max}.}
        \begin{tabular}{l|l|l}
    \toprule
        Value & Cluster 0 & Cluster 1 \\
        \midrule
        62 &  2'363 &    0 \\
        63 & 19'710 &    0 \\
        64 & 16'566 &    0 \\
        66 &      3 & 3233 \\
    \bottomrule
    \end{tabular}
    \label{tab:TTLmax_count_1BD_feature}
\end{table}

\paragraph{Eliminating the backdoor}
To eliminate the backdoor, one could remove weights frequently used by the backdoor feature whenever the trigger is used.\footnote{See discussion in section \ref{sec:RemoveBD}.} However, considering all active paths is generally infeasible, as both clusters might use all paths. Instead, we focus on the most typical paths used by the clusters --- specifically those used more than $50$ times by a cluster.\footnote{Algorithm \ref{alg:CompareAPBetweenDatasets} in the Appendix details the elimination algorithm. Setting $T=50$ will only return paths used more than $50$ times.} 
Removing weights associated with Cluster 1 that originates from \texttt{TTL\_max} yields the results shown in \autoref{tab:feat_bd_restored} (Model after elimination). Compared to the backdoored model in the same table, the backdoor behaviour have largely been eliminated without significantly degrading the model's predictive behaviour, achieving this in a cost-efficient manner.

\begin{table}[h]
\centering
\caption{Accuracy of model (before and after elimination)}
\label{tab:feat_bd_restored}
\begingroup
\renewcommand{\arraystretch}{1.12} % 15% taller rows; tweak 1.10–1.25 as needed
\begin{tabular}{c c l|cc|cc}
\toprule
& & & \multicolumn{2}{c|}{\textbf{Backdoored model}} & \multicolumn{2}{c}{\textbf{Model after elimination}} \\
\cmidrule(lr){4-7}
& & & Clean & Poisoned & Clean & Poisoned \\
\midrule
\multirow{3}{*}{\rotatebox[origin=c]{90}{1 feature}}
& & All data  & 99.29\% & 97.79\%           & 99.30\% & 98.72\% \\
& & Benign    & 99.48\% & 99.98\%           & 99.50\% & 98.90\% \\
& & Malicious & 91.34\% & \textbf{5.19\%}   & 90.91\% & \textbf{90.91\%} \\
\midrule
\multirow{3}{*}{\rotatebox[origin=c]{90}{2 features}}
& & All data & 99.37\% & 97.68\% & 99.51\% & 99.74\% \\
& & Benign & 99.57\% & 99.99\% & 99.71\% & 99.96\% \\
& & Malicious & 90.91\% & \textbf{0.00\%} & 90.91\% & \textbf{90.48\%} \\
\bottomrule
\end{tabular}
\endgroup
\end{table}

\subsection{Experiment 2: Backdoor Using Two Features}\label{sec:TwoBDFeaturesExperiment}

Our second experiment has a similar setup as the first one. Here, the backdoor is implemented using two backdoor features: \texttt{TTL\_max}
and \texttt{TTL\_min}. As well as setting \texttt{TTL\_max} to $66$, the trigger also uses a value of $61$ for \texttt{TTL\_min}. This backdoor was also successfully implemented, with an accuracy of $99.23\%$ on clean data and a poison accuracy of $99.98\%$. 

\begin{figure}\centering
\includegraphics[width=0.375\textwidth]{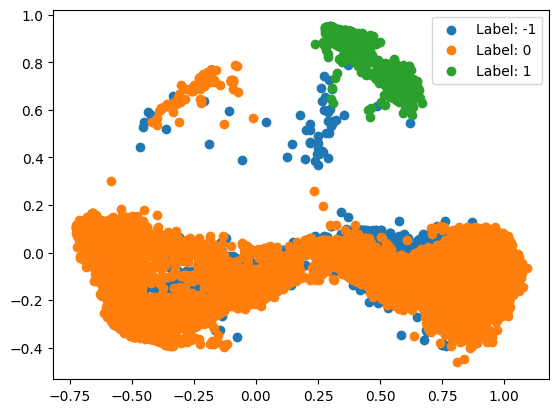} 
\caption{Clustering of feature contributions for all benign predictions having two backdoor features.}
\label{fig:contribution_clustering_2_bd}
\end{figure}

\begin{figure}
        \centering
        \includegraphics[width=0.375\textwidth]{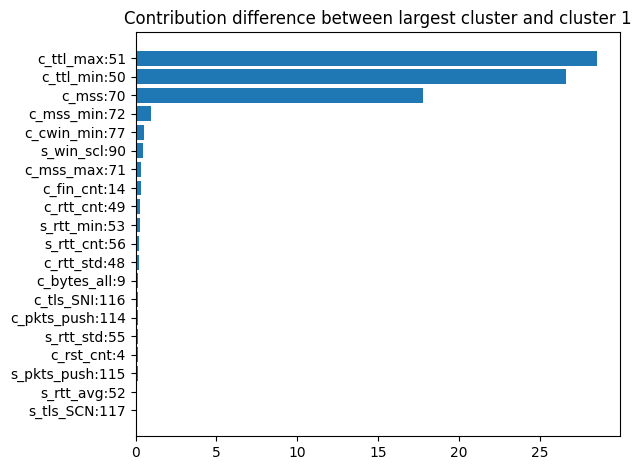}
        \caption{Contribution difference between cluster 0 and 1 in mean square difference.}
        \label{fig:contri_diff_2_bd}
\end{figure}

\begin{figure}
        \centering
        \includegraphics[width=0.375\textwidth]{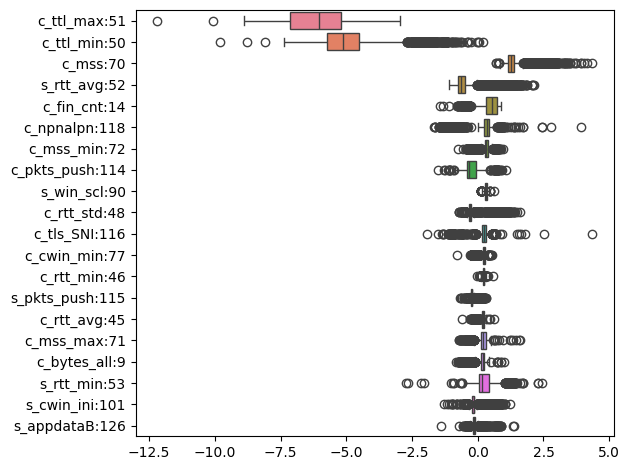}
        \caption{Feature contributions for cluster 1.}
        \label{fig:cont_cluster1_2_bd}
\end{figure}

\paragraph{Detecting the backdoor}
We use the same analysis as in the first experiment. As seen in \autoref{fig:contribution_clustering_2_bd}, we again get two clusters. Although not as distinct as in the first experiment, Cluster 0 and Cluster 1 are still clearly separable. The contribution differences in \autoref{fig:contri_diff_2_bd} show that \texttt{TTL\_max} and \texttt{TTL\_min} differ the most between the clusters. A closer inspection of the feature contributions of Cluster 1 (see \autoref{fig:cont_cluster1_2_bd})
reveals that these two features are the main contributors for predicting benign behaviour.
Additionally, \autoref{tab:TTL_min_max_count_2BD_feature} show that Cluster 1 only uses a single value for both features, indicating that the model will behave differently when these values are present. 
To assess their impact on the model, we injected clean data with both values. The results, shown in the lower part of \autoref{tab:feat_bd_restored} (Backdoored model), demonstrates a substantial drop in predictive performance for the malicious class, as seen under the `Poisoned' column. This strongly suggests that the model is
prone to predict benign behaviour whenever these values are present, confirming their role as backdoor triggers.
% that these are backdoor triggers injected into the model. 

\paragraph{Eliminating the backdoor}
We use the same technique as in the first experiment to eliminate the backdoor, where weights associated with Cluster 1 that originates from \texttt{TTL\_max} and \texttt{TTL\_min} are set to zero. The results from \autoref{tab:feat_bd_restored} (Model after elimination; bottom part) shows that 
the backdoor trigger is no longer effective, and the accuracies are roughly the same before and after eliminating weights on the clean data.

\begin{table}[h]
\centering
\caption{Frequency of \texttt{TTL\_min} and \texttt{TTL\_max}.}
\label{tab:TTL_min_max_count_2BD_feature}
\begingroup
\renewcommand{\arraystretch}{.9} % 15% taller rows; tweak 1.10–1.25 as needed
\begin{tabular}{l|cc|cc}
\toprule
& \multicolumn{2}{c|}{\textbf{\texttt{TTL\_min}}} & \multicolumn{2}{c}{\textbf{\texttt{TTL\_max}}} \\
\cmidrule(lr){2-5}
& Cluster 0 & Cluster 1 & Cluster 0 & Cluster 1 \\
\midrule
\multirow{3}{*}
\text{61}  &    430 & 3'143 &      0 &     0 \\
62         &  2'089 &     0 &  2'089 &     0 \\
63         & 19'314 &     0 & 19'314 &     0 \\
64         & 16'424 &     0 & 16'424 &     0 \\
66         &      0 &     0 &    430 & 3'143 \\
\bottomrule
\end{tabular}
\endgroup
\end{table}

\section{Discussion and Related Work}\label{sec:discussion}

Our experiments demonstrate that backdoors are a potent attack in the IDS setting, with successful backdoors requiring only $1\%$ of the training data to be poisoned\footnote{We note, however, that the backdoor percentage may vary depending on data distributions and model architectures~\cite{wang2024demystify}.}. Furthermore, we have proposed two different backdoor mitigation techniques: removal of backdoors and alerting on backdoor-like behaviour. Which of these techniques is most appropriate will depend on deployment-specific requirements.

Comparing the contribution of one backdoored feature in the first experiment (\autoref{fig:contri_diff_1_bd}) with multiple backdoored features in the second experiment (\autoref{fig:contri_diff_2_bd}), we see that the explanatory contributions are reduced when using two features, indicating that our approach might be less robust against triggers that use multiple features. We note, however, that this is based on a single, synthetic dataset and may be an artefact of the backdoors investigated. 

While our approach shows promise, further experiments are necessary to show that it generalises beyond this dataset and setting. For instance, a limitation of our experiments is that both the backdoor insertion and model training were performed by us. Future work could include scenarios where the backdoor is implanted by an external party, as this would better reflect real-world conditions. Moreover, comparisons with other backdoor detection and removal methods on the same dataset should also be conducted.

Our detection technique depends on the availability of data where the trigger is present. This may not be possible in specific settings, such as when using public models\footnote{E.g. it is common to use models available on sites such as Huggingface.} or in a federated setting. However, an advantage of our approach is that it is solely based on active paths and local contributions, and thus does not require access to a non-poisoned dataset, unlike other methods~\cite{wang2019neural,xu2021detecting,wu2021adversarial}.

Our approach cannot distinguish between backdoors and strong overfitting or feature correlations. This requires the end-user of the technique to possess sufficient domain knowledge. 
For the area of intrusion detection, analysts must recognise ``anomalous behaviours", such as a model that predicts solely based on TTL-values. However, our method provides inherent explanations that support this analysis. Many of these limitations are inherent in other backdoor detection techniques. Our approach is limited to piecewise linear activations (ReLU, Leaky ReLU) and requires identifiable and distinct active paths for elimination, though extension to convolutional architectures seems feasible.

The closest related work to our detection method, \emph{activation clustering}\cite{chen2018detecting}, clusters similar observations, but only considers final-layer activations. As a consequence, they lose feature explanation and require retraining for backdoor elimination. \emph{BadActs} \cite{yi2024badacts} also resembles our method, as it compares activation differences within the network to detect backdoors. However, this method detects anomalies by assuming the activation space adheres to a Gaussian distribution, and like activation clustering, it does not provide explainability.

Our backdoor elimination technique only requires computing and comparing layer activations, which can be done via a single forward pass. Thus, there is no need to retrain the model (which is the case for e.g. BadActs \cite{yi2024badacts}), reducing computation overhead and falling under the general category of \emph{model editing}\footnote{\emph{Model editing} means that model weights are directly changed.} \cite{wang24LLMknowedit,yang2025mirage,hoyheim2025explainable}. Recently, these methods have focused heavily on modifying large language models \cite{wang24LLMknowedit,meng2022locating} to update factual associations. Subsequent claims suggest that the performance loss is substantial \cite{yang2025mirage}, affecting other inputs such as clean samples. This will require further investigation in the IDS setting. Finally, we note that while previous work on \emph{active paths} \cite{hoyheim2025explainable} --- 
 which we base our work on\footnote{See section \ref{sec:localCont_activePath} for details.} --- only removes paths that do not contribute, our method also eliminates paths that significantly contribute to predictions. This is a novel usage of active paths.

\section{Conclusion}\label{sec:conclusion}

From the observation that backdoor triggers in machine learning models are often manifested in abnormally strong paths during forward propagation in a neural network, we have presented a novel approach that exploit this to detect possible backdoors. The approach is explainable by design and can be used to remove backdoors in a resource-efficient manner directly. Crucially, this is achieved without the need to retrain the model and/or relabel the training data, both of which can be very cost-intensive.

We have demonstrated our approach in an intrusion detection context, where one could either remove the backdoor from the model, or choose to keep it and instead explore the explainability aspect of our approach by alerting on backdoor-like behaviour. Further work will focus on developing stronger experimental evidence, including comparisons and contrasts with other techniques using the same dataset.

\section*{Appendix: Algorithms for Eliminating Backdoors}\label{appendix:RemoveBackdoor}

Algorithm \ref{alg:CountAPs} computes the frequency with which each weight is used in an active path. A path is only considered if it has been utilised  more than $T$ times, ensuring that only the most frequently used paths are considered.

\begin{algorithm}
\caption{Count weights in active paths (CWAP)}\label{alg:CountAPs}
\begin{algorithmic}[1]
\Require A trained sequential model $\mathcal{M}$, an input dataset $D$, and a minimum usage threshold $T$.
\Ensure Count of weights in active paths
\State Compute layer-wise activations for all samples in $D$ using $\mathcal{M}$, store non-zero activations for each sample in $A$
\State Let $N$ be the number of samples in $D$
\State Let $W$ be all weights in $\mathcal{M}$ 
\State Initialize dictionary $\texttt{all\_active\_paths}$ 
\For{each sample index i from 1 to $N$}
    \State $\texttt{all\_active\_paths}[i]$ $\gets$ $\mathbbm{1}$ $[W $ connected to node in $ A]$ 
\EndFor
\State Count unique paths in $\texttt{all\_active\_paths}$, store paths and counts in $\texttt{active\_path\_count}$
\State Initialize $W_{\text{count}}$ with zeros in the shape of $W$
\For{ each ($\text{path, count}$) in $\texttt{active\_path\_count}$}
    \If{count $> T$}
        \State $W_{\text{count}} \gets W_{\text{count}} + \text{path}$
    \EndIf
\EndFor

\State \Return $W_{\text{count}}$

\end{algorithmic}
\end{algorithm}

Algorithm \ref{alg:CompareAPBetweenDatasets} compares commonly used active paths between two datasets and reports differences in the weight matrix connecting the input layer to the first hidden layer. When applied to backdoored versus clean data, the algorithm identifies activation differences, indicating whether a path is unique to one dataset ($-1$ or $1$), or used by both or neither of them ($0$).

\begin{algorithm}
\caption{Compare Active Paths Between Two Datasets}\label{alg:CompareAPBetweenDatasets}
\begin{algorithmic}[1]
\Require A trained sequential model $\mathcal{M}$, two datasets $D_1$ and $D_2$, and a minimum usage threshold $T$
\Ensure Difference in activation usage and usage weights for each dataset

\State Get weight usage for $D_1, D_2$ with \textsc{CWAP}, save in $W_1, W_2$

\State Compute row-wise sums of $W_1[0], W_2[0]$ \Comment{The first instance are weights between the input and first hidden layer}
\State Convert sums to binary indicators: 1 if sum $> 0$, else 0
\State Compute difference: $\text{usage\_diff} \gets \text{indicator}_1 - \text{indicator}_2$

\State \Return $\text{usage\_diff}, W_1, W_2$

\end{algorithmic}
\end{algorithm}

\bibliographystyle{plain}

\end{document}